\documentclass[aps,prl,twocolumn,groupedaddress,showpacs,amsmath,amssymb]{revtex4}
\usepackage{graphicx}

\begin{document}


\title{Formation of ultracold polar molecules in the rovibrational ground state}

\author{J. Deiglmayr}
\author{A. Grochola}
\altaffiliation{Also at the Institute of Experimental Physics,
Warsaw University, Poland}
\author{M. Repp}
\author{K. M\"ortlbauer}
\author{C. Gl\"uck}
\author{J. Lange}
\author{O. Dulieu}
\altaffiliation{Permanent address: Laboratoire Aim\'e Cotton, CNRS,
Universit\'e Paris Sud XI, Orsay, France}
\author{R. Wester}
\author{M. Weidem\"uller}

\email{weidemueller@physik.uni-freiburg.de}

\affiliation{Albert-Ludwigs-Universit\"at Freiburg, Physikalisches
Institut, Hermann-Herder-Str. 3, 79104 Freiburg, Germany}

\date{\today}

\begin{abstract}
Ultracold LiCs molecules in the absolute ground state X$^1\Sigma^+$,
$v''$=0, $J''$=0 are formed via a single photo-association step
starting from laser-cooled atoms. The selective production of
$v''$=0, $J''$=2 molecules with a 50-fold higher rate is also
demonstrated. The rotational and vibrational state of the ground
state molecules is determined in a setup combining depletion
spectroscopy with resonant-enhanced multi-photon ionization
time-of-flight spectroscopy. Using the determined production rate of
up to $5\times10^3$ molecules/s, we describe a simple scheme which can
provide large samples of externally and internally cold dipolar
molecules.
\end{abstract}



\maketitle


A gaseous cloud of translationally ultracold molecules, i.e. well
below a temperature of 1\,mK, in their rovibrational ground state is
the starting point for many intriguing scientific applications, such
as the exploration of quantum phases in dipolar gases
\cite{micheli2006, pupillo2008}, the development of quantum
computation techniques \cite{rabl2006}, precision measurements of
fundamental constants \cite{zelevinsky2008}, and the investigation
and control of ultracold chemical reactions \cite{tscherbul2006}. A
number of experimental approaches are currently being studied to
prepare and manipulate ultracold molecules
\cite{doyle2004,dulieu2006}. Up to now, the formation of ultracold
molecules in the lowest vibrational level of the electronic ground
state has been demonstrated for ultracold KRb~\cite{nikolov2000},
RbCs~\cite{sage2005}, and Cs$_2$~\cite{viteau2008}, in all cases
using complex photo-association schemes in an ultracold gas of atoms
involving both continuous and pulsed laser fields. In an alternative
approach, ultracold gaseous samples of magneto-associated, weakly
bound molecules, so-called Feshbach molecules, have recently been
transferred into deeply bound vibrational states, yet not the lowest
state, by coherent adiabatic passage
\cite{winkler2007,ospelkaus2008,danzl2008} .

\begin{figure}
\includegraphics[width=\columnwidth,clip]{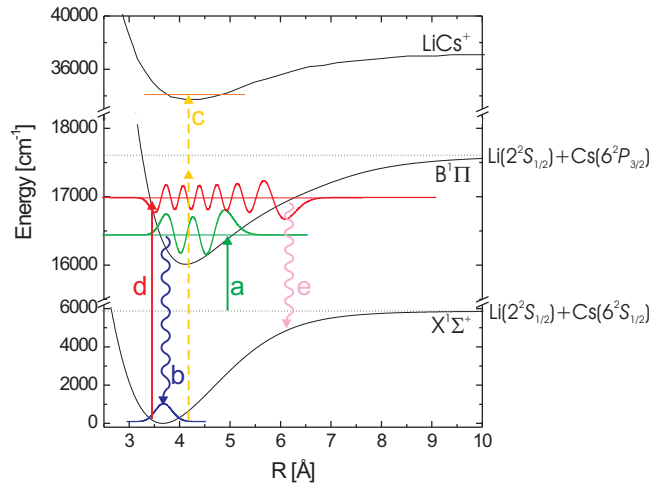}
\caption{\label{fig:scheme} Sketch of the excitation and detection
scheme. a) photo-association by Ti:Sa laser at 946nm, b) spontaneous
decay into deeply bound ground state molecules, c) two photon
ionization with resonant intermediate state. For depletion
spectroscopy: d) excitation of ground state molecules,  e)
redistribution of ground state population; potentials curves from
Ref.~\cite{staanum2007,stein2008,korek2006}.}
\end{figure}

In this work we demonstrate the production of ultracold LiCs
molecules in the absolute vibrational \emph{and} rotational ground
state X$^1\Sigma^+$,$v''$=0,$J''$=0, by scattering the light of a
single narrow-band photo-association (PA) laser off pairs of
magneto-optically trapped lithium and cesium atoms. We unambiguously
assign the produced quantum state using high resolution,
rotationally selective depletion spectroscopy combined with
resonance enhanced multi-photon ionization time-of-flight
(REMPI-TOF) mass spectrometry. The sequence of the formation and
detection steps is schematically shown in Fig.~\ref{fig:scheme}.

$^{133}$Cs and $^{7}$Li atoms out of a double species oven are
slowed in a Zeeman slower and trapped in an overlapped
magneto-optical trap (MOT) for lithium and a forced dark-spot MOT
(SPOT)~\cite{ketterle1993} for cesium. We trap 4$\times$10$^7$
cesium atoms and 10$^8$ lithium atoms at densities of
3$\times$10$^{9}~$cm$^{-3}$ and 10$^{10}~$cm$^{-3}$ respectively.
Time-of-flight expansion was used to measure a cesium temperature of
250(50)$\mu$K. Due to the large photon recoil and unresolved
hyperfine structure of the excited state, the lithium atoms have a
temperature of hundreds of $\mu$K. In the cesium SPOT, 97\% of the
atoms are in the lower hyper-fine ground state $F$=3, while in the
lithium MOT 80\% of the atoms are in the upper hyper-fine state
$F$=2. Therefore the atoms collide mainly on the
Li(2$^2S_{1/2},F$=2)+Cs(6$^2S_{1/2},F$=3) asymptote. For PA 500mW of
light from a Ti:Sa laser are collimated to a waist of 1.0\,mm and
passed through the center of the trap region, left on continuously
during all measurements. For the detection of formed molecules,
first all cesium atoms in the trap are quenched to the ground state
by blocking the cesium repumper, because two photon ionization of
excited cesium atoms would form a strong background signal for the
detection of LiCs$^+$ ions. After 0.6\,ms a pulsed dye laser (pulse
energy 4\,mJ, pulse length 7\,ns, bandwidth 2\,GHz) ionizes ground
state molecules. The laser beam is collimated to a waist of
$\sim$\,5\,mm and is aligned to pass roughly one beam diameter below
the trapped atoms in order to prevent excessive ionization of atoms.
The ions are then detected in a high resolution time-of-flight mass
spectrometer and are counted in a single-ion counting setup (for
details see Ref.~\cite{kraft2006,kraft2007}). The extraction fields
of 47\,V/cm are switched on only 0.5\,ms before the ionization pulse
for 1\,ms, so that the PA is performed mainly under field-free
conditions. This experimental cycle is repeated at 20\,Hz.

We determine the relative collision energy of lithium and cesium
atoms by fitting the shape of a narrow, temperature-broadened PA
resonance with the model of Ref.~\cite{jones1999}. Assuming a
natural linewidth of $\gamma$=7\,MHz, we deduce a relative collision
temperature of 530(80)$\mu$K, dominated by the temperature of the
lithium atoms. This is well below the Li-Cs $p$-wave centrifugal
barrier of 1.6\,mK derived from the C$_6$ dispersion coefficient of
Ref.~\cite{marinescu1994}. Assuming a Boltzmann distribution of
collision energies, we expect a $p$-wave contribution on the order
of 5\% and no contributions from higher partial waves.

For the production of ground state molecules, photo-association is
performed via the B$^1\Pi$ state correlated to the
Li(2$^2S_{1/2}$)+Cs(6$^2P_{3/2}$) asymptote. We identify vibrational
levels from $v'$=35 just below the asymptote down to $v'$=4 and find
excellent agreement with the energies calculated from an
experimental energy curve for this state~\cite{stein2008}.

\begin{figure}[t]
\includegraphics[width=0.95\columnwidth,clip]{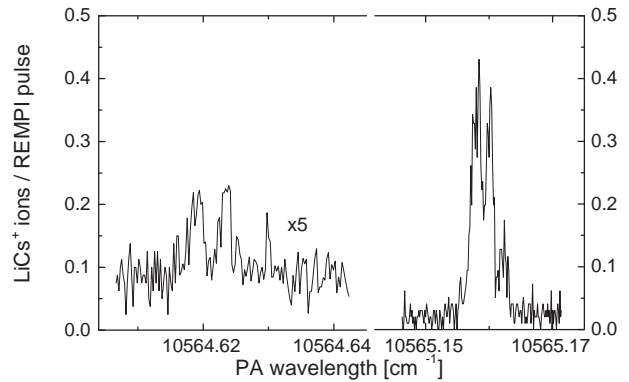}
\caption{\label{fig:palines}The $v'$=4,$J'$=1 (left trace) and
$J'$=2 (right trace) PA resonances in the B$^1\Pi$ state. The $J'$=1
trace has been enlarged by a factor of 5 for better visibility. }
\end{figure}

For the experiments presented in this work we focus on the
$v'$=4,$J'$=1 and $J'$=2 levels of the B$^1\Pi$ state which are
addressed with PA light of 946.56\,nm. Typical PA scans of the
corresponding resonances are shown in Fig.~\ref{fig:palines}. Both
resonances show substructure which reflects the molecular hyperfine
interactions. For PA we always choose the strongest component. It is
noteworthy to mention the strongly increased PA rate for the $J'$=2
over the $J'$=1 resonance by roughly a factor of 20. This is in
contrast to ratios $\sim$1 observed in our experiment for higher
vibrational levels $v'>20$. A full analysis of the observed PA
lineshapes will be subject of further studies.

The excited B$^1\Pi$, $v'$=4 molecules decay spontaneously only into
the X$^1\Sigma^+$ state~\cite{stein2008}.
Using the experimental potential curves from
Ref.~\cite{stein2008,staanum2007} and an ab-initio $R$-dependent
dipole moment function~\cite{foonoteDipolemomentFunction}, Einstein
$A$ coefficients for the spontaneous decay from B$^1\Pi$,$v'$=4 into
X$^1\Sigma^+$ levels are calculated. Tab.~\ref{tab:grounddist} shows
the deduced relative population of the X$^1\Sigma^+$ levels. From
these calculations 23\% of all excited state molecules are expected
to decay into the X$^1\Sigma^+$,$v''$=0 level. We note that nearly
all excited $v'$=4 molecules are expected to decay into bound
molecules, since the sum over all Franck-Condon (FC) factors for
decay into the X$^1\Sigma^+$ state is close to
unity~\cite{drag2000}.

\begin{table}[b]
\begin{tabular}{|r|ccccccccccccc|}\hline
X$^1\Sigma^+$,$v''$= & \textbf{0}& 1&  2& 3& 4& 5& 6& 7& 8&
9&10&$\Sigma$(11-20) & $\Sigma$($>$20)\\ \hline
Rel. pop. [\%] & \textbf{23}& 1& 12& 2& 3& 8& 2& 1& 5& 6& 2& 35& 0 \\
\hline
\end{tabular}
\caption{\label{tab:grounddist} Calculated relative population of
vibrational states in the X$^1\Sigma^+$ state after
photo-association via B$^1\Pi$,$v'$=4.}
\end{table}

\begin{figure}[t]
\includegraphics[width=\columnwidth,clip]{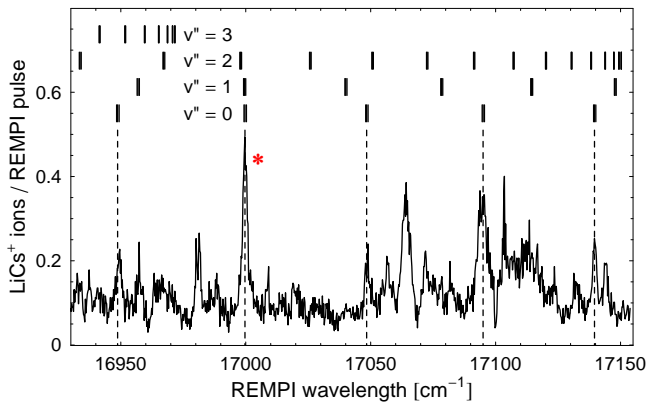}
\caption{\label{fig:rempilines}Resonant-enhanced multi-photon
ionization of ground state molecules produced by photo-association
via B$^1\Pi$,$v'$=4,$J'$=2. In the upper part, calculated line
positions between all vibrational levels in the X$^1\Sigma^+$ and in
the B$^1\Pi$ state are marked. The red star (*) marks the resonance
used in the depletion spectroscopy. At the moment we do not have a
full assignment of all observed resonances.}
\end{figure}

The formation of X$^1\Sigma^+$ molecules is detected by two-photon
ionization with typical wavelengths in the range of 575\,nm to
600\,nm. At these energies, vibrational levels $v''$=0-4 of the
X$^1\Sigma^+$ state can be ionized via the well known intermediate
B$^1\Pi$ state. Fig.~\ref{fig:rempilines} shows a REMPI scan
together with expected positions for transitions from the
X$^1\Sigma^+$,$v''$=0-3 levels to intermediate levels in the
B$^1\Pi$ state. Resonances are clearly visible at the expected
positions for transitions from X$^1\Sigma^+$,$v''$=0 to intermediate
levels B$^1\Pi$,$v'$=13-17. However some of these resonances could
also include contributions from higher vibrational states. The
intensities required for the ionization of ground state molecules
strongly saturate the first resonant bound-bound transitions,
therefore the rotational structure of the ground state levels is not
resolved.

In order to further identify the internal quantum states of the
ultracold ground state molecules we perform depletion spectroscopy
of the formed ground state molecules~\cite{wang2007}. An additional
narrow band laser optically pumps population out of rovibrational
levels of the X$^1\Sigma^+$ ground state (Fig.~\ref{fig:scheme}).
Those levels are coupled to specific rovibrational levels in the
B$^1\Pi$ potential from which spontaneous decay leads to
higher-lying vibrational levels. The full scheme for the depletion
spectroscopy is as follows: with the PA laser locked at a chosen
resonance, the REMPI laser is set to selectively ionize one
vibrational ground state and a cw dye laser (bandwidth $\sim$5\,MHz,
typically 40mW, $\omega_0$=0.7\,mm), aligned collinear with the PA
light, is scanned. Rotational components of the chosen vibrational
level are detected as a reduction of the ion count rate when the
narrow cw dye laser is resonant with transitions to excited state
levels. The expected positions of the depletion resonances are given
by $\hbar\omega_0+B'J'(J'+1)-B''J''(J''+1)$ where $\hbar \omega_0$,
the term energy difference between excited and ground state
vibrational level, and $B'$ ($B''$), the excited state (ground
state) rotational constant, are calculated from experimental
potential energy curves \cite{staanum2007,stein2008}.

\begin{figure}[b]
\includegraphics[width=0.85\columnwidth,clip]{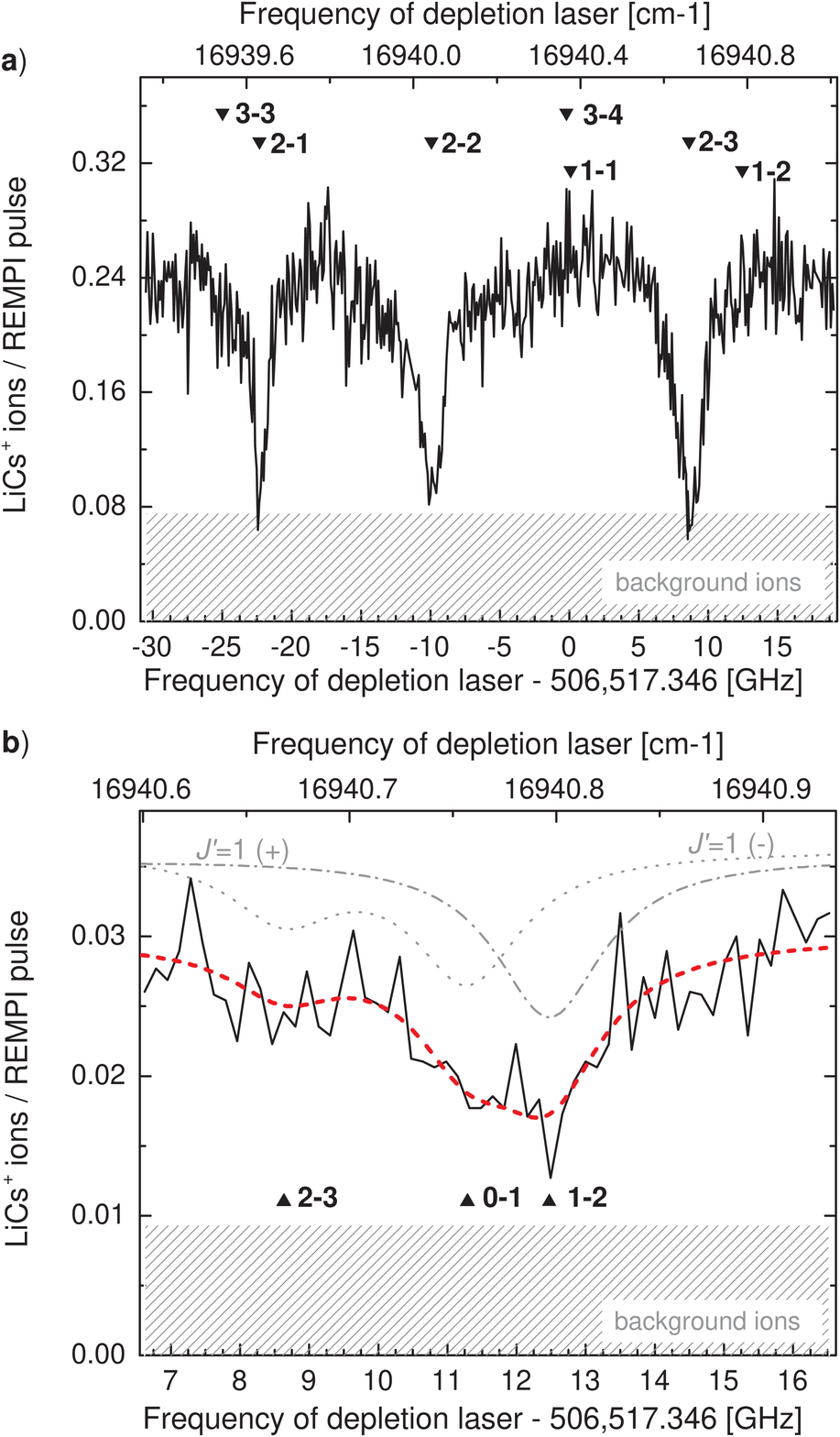}
\caption{\label{fig:depletion}Depletion laser scan with the PA laser
resonant to the B$^1\Pi$,$v'$=4,$J'$=2 level (a) and to the
$v'$=4,$J'$=1 level (b). In the upper part of the graphs, calculated
wavelengths for transitions from X$^1\Sigma^+$,$v''$=0,$J''$ to
B$^1\Pi$,$v'$=12,$J'$ (labeled \mbox{$\textsf{J''}$-$\textsf{J'}$})
are marked. The dotted and dash-dotted grey lines in (b) are fitted
spectra (see text for details) assuming PA via the (+) or (-) parity
component of  J'=1 respectively, offset for visibility. The dashed
(red) line is a fit combining both parities. The background at large
formation rates as in (a) is dominated by LiCs$^+$ ions from
off-resonant excitation of other vibrational ground state levels.
For small formation rates as in (b), spurious detection of fast
Cs$^+$ ions constitutes the main background~\cite{kraft2007}.}
\end{figure}

Depletion scans were performed for
X$^1\Sigma^+$,$v''$=0 molecules, ionized via the intermediate
B$^1\Pi$,$v'$=14 level at an ionization wavelength of
16999.4\,cm$^{-1}$ (marked with a star in
Fig.~\ref{fig:rempilines}). In the scan of
Fig.~\ref{fig:depletion}\,a), the molecules are produced by PA via
B$^1\Pi$,$v'$=4,$J'$=2 (Fig.~\ref{fig:palines}, right trace). We
observe that excitation on the transitions from
X$^1\Sigma^+$,$v''$=0,$J''$=2 to
B$^1\Pi$,$v'$=12,$J'$=1-3 reduces the ion count rate down to
the background level. Therefore, the ions at this REMPI resonance originate
predominantly in the X$^1\Sigma^+$,$v''$=0 level. The
spontaneous decay of the B$^1\Pi$,$v'$=4,$J'$=2 level occurs
only via the Q-branch ($\vartriangle$$J$=0) leading to the
population of only $J''$=2 rotational levels, as shown by the
rotational assignment in
Fig.~\ref{fig:depletion}\,a)~\cite{foonoteJ2}. From the measured
spectrum we derive the excited state rotational constant
$B''$=3.10(5)\,GHz in excellent agreement with the calculated value
of 3.096\,GHz. In combination with the calculated ground state
rotational constant $B''$ one gets $\hbar
\omega_0$=16895.75(2)\,cm$^{-1}$ which is also in very good
agreement with the expected value of 16895.77\,cm$^{-1}$. The
depletion resonances show a width of about 2\,GHz, which can be
attributed to hyperfine broadening of the excited states (typically
500\,MHz) and strong saturation of the transition. Hyperfine
structure of deeply bound ground state levels is expected to be
below 1\,MHz and is therefore not resolved in the current
experiment.

Fig.~\ref{fig:depletion}\,b) shows a depletion scan for $v''$=0
molecules produced via the PA resonance B$^1\Pi$,$v'$=4,$J'$=1
(Fig.~\ref{fig:palines}, left trace). With $\hbar\omega_0$ and $B'$
from the depletion spectrum of Fig.~\ref{fig:depletion}\,a) and the
calculated value for $B''$ the exact positions of the
depletion resonances are known. Using the widths found for the $J'$=2
depletion resonances and relative population strengths from
H\"onl-London-Factors, we fit spectra for different excited state
parities to the data in Fig.~\ref{fig:depletion}\,b). The only
free fit parameters are the level of the undepleted ion signal and the
depletion depth. The precisely known positions of the depletion
resonances make a single excited state parity unlikely and we
attribute the observed spectrum to PA via both parity components
with equal strength leading to population of
X$^1\Sigma^+$,$v''$=0,$J''$=0,1 and 2 states~\cite{foonoteJ1}.

For molecules in the absolute ground state
X$^1\Sigma^+$,$v''$=0,$J''$=0, produced by PA via
B$^1\Pi$,$v'$=4,$J'$=1, on average 5$\times$10$^{-3}$~LiCs$^+$ ions
are detected per ionization pulse. Since the experiment is running
at 20\,Hz this yields a detection rate of 0.1\,ions/s. Taking into
account a geometric overlap factor of 40\%, a typical on-resonance
ionization probability of 1\%, and a detector efficiency of
20\%~\cite{fraser2002}, this results in a production rate of about
1$\times$10$^2$\,molecules/s in the $v''$=0,$J''$=0 absolute ground
state. For molecules in the X$^1\Sigma^+$,$v''$=0,$J''$=2 state,
produced by PA via B$^1\Pi$,$v'$=4,$J'$=2, on average 0.2\,LiCs$^+$
ions per ionization pulse are detected. Following the same argument
as above, this results in a production rate of about
5$\times$10$^3$\,molecules/s in the ground state level $v''$=0,
comparable to the rate given in Ref.~\cite{sage2005} for RbCs and
roughly a factor 20 smaller than the rate given in
Ref.~\cite{viteau2008} for the homonuclear Cs$_2$. However, we want
to point out that we determine rates for the population of a single
vibrational and rotational state, while in the cited references the
produced molecules are distributed over a range of rotational
states. The translational temperature of the molecules is estimated
based on the atomic values and a simple kinematic model to be about
260\,$\mu$K, dominated by the temperature of the heavier cesium;
contributions from photon recoil during the single absorption and
emission cycle are not significant.

As the presented scheme for the formation of molecules relies on
spontaneous decay, it allows the continuous accumulation of
molecules in the absolute ground state. Due to the large vibrational
level spacing in LiCs accidental re-excitation of formed molecules
by the narrow-band PA laser is very unlikely. This accumulative
approach is in contrast to schemes involving adiabatic transfer. The
molecules produced in the presented setup could be trapped in an
electrostatic trap~\cite{kleinert2007}. Alternatively, one could start from a
mixture of lithium and cesium atoms in an optical dipole
trap~\cite{mudrich2002}, where typically 10-fold increased densities
could lead to a production rate on the order of
5$\times$10$^5$\,molecules/s for ground state molecules in
$v''$=0,$J''$=2. On a time scale of seconds, a large part of
the trapped atom pairs may therefore be transformed into molecules.
In a final step of adiabatic transfer using microwaves at 22.5 and
11.2\,GHz, the $J''$=2 molecules could be transferred to $J''$=0.
Excited vibrational states would be constantly removed by inelastic
collisions with ultracold atoms, as shown for
Cs$_2$+Cs~\cite{staanum2006,zahzam2006}. Sympathetic cooling in the
dipole trap~\cite{mudrich2002} may yield molecular samples with
temperatures down to the nanokelvin regime. This may open up am efficient route to form stable quantum gases of molecules in their
absolute internal ground state.

\begin{acknowledgments}
We thank S.D. Kraft and P. Staanum for contributions at the early
stage of the experiment. We also thank E. Tiemann and A. Pashov for
providing experimental LiCs potentials before publication and,
together with J. Hutson, for fruitful discussions. This work is
supported by the DFG under WE2661/6-1 in the framework of the
Collaborative Research Project QuDipMol within the ESF EUROCORES
EuroQUAM program. JD acknowledges partial support of the
French-German University. AG is a postdoctoral fellow of the
Alexander von Humboldt-Foundation.
\end{acknowledgments}

\bibliography{mixtures-short}

\begin{thebibliography}{32}
\bibitem[{foo({\natexlab{c}})}]{foonoteJ1}
\bibinfo{note}{The strong $d$-wave contribution to the PA via
  B$^1\Pi$,$v'$=4,$J'$=2 couples also to the negative parity component of
  B$^1\Pi$,$v'$=4,$J'$=1. Decay from this component populates
  X$^1\Sigma$,$v''$=0,$J''$=0,2 as observed. The additional population of
  $J''$=1 can be explained by p-wave contributions to the continuum wave
  function.}
\end{thebibliography}

\end{document}